\documentclass[final]{aipproc}
\layoutstyle{6x9}

\def\ga{\gamma}
\def\de{\delta}
\def\si{\sigma}
\def\ch{\chi}
\def\om{\omega}
\def\De{\Delta}
\def\fr#1#2{{{#1} \over {#2}}}
\def\hydrogen{$H$}
\def\antihydrogen{$\overline{H}$}
\def\h{\hydrogen}
\def\ah{\antihydrogen}
\def\ket#1{|{#1}\rangle}
\def\half{{\textstyle{1\over 2}}}
\newcommand{\beq}{\begin{equation}}
\newcommand{\eeq}{\end{equation}}
\newcommand{\bea}{\begin{eqnarray}}
\newcommand{\eea}{\end{eqnarray}}

\begin{document}

\title{Tests of Lorentz Symmetry with Penning Traps and Antihydrogen
\footnote{Proceedings for invited talk at Eighth International Conference on Low Energy Antiproton Physics (LEAP '05),
Bonn, Germany, 16-22 May 2005.}}


\author{Neil Russell}{
  address={Physics Department, Northern Michigan University, Marquette, MI 49855, USA}
}

\begin{abstract}
Possibilities for testing Lorentz symmetry using precision experiments
with antiprotons in Penning traps
and with antihydrogen spectroscopy are reviewed.
Estimates of bounds on relevant coefficients for
Lorentz violation in the Standard-Model Extension (SME)
are considered.
\end{abstract}

\maketitle

\section{The Standard-Model Extension}
Einstein's General Relativity and the Standard Model of particle physics
provide an excellent description of nature.
They are expected to merge into a single theory
at energies near the Planck scale of $10^{19}$~GeV.
Experiments cannot access these energies,
but one can at least hope to see suppressed effects
of Planck-scale physics in suitable high-precision experiments.
One possible signal of an underlying unified quantum gravity theory
is the violation of Lorentz symmetry.
In this proceedings of the
2005 International Conference on Low-Energy Antiproton Physics (LEAP 05),
some possibilities for detecting Lorentz and CPT violation
in experiments with low-energy antiprotons will be reviewed.
Most of the results summarized here are presented in more detail in
the published literature, and a few speculative ideas are discussed.

Any effects of Lorentz violation can be described
using effective field theory \cite{kp:95}.
To be realistic, both General Relativity
and the Standard Model must be included
together with coordinate-independent Lorentz violation.
These components make up the effective field theory
called the Standard-Model Extension (SME).
It provides a general description of all possible realistic
Lorentz violations in nature.
The Minkowski limit of this theory
has been studied for more than a decade
\cite{ck:sme,CPT:04},
and recent work
\cite{grav}
has clarified the key ideas in curved spacetime.
Among the underlying ideas are string-theory concepts
and various types of symmetry breaking \cite{strings}.
For some other perspectives on Lorentz violation
presented at LEAP '05, see \cite{nm}.

The focus in this article is primarily on
theoretical and experimental studies of
Lorentz tests investigating the SME with
Penning traps
\cite{ptth, ptexp,ptexp2}
and with spectroscopy of
hydrogen (\h) and antihydrogen (\ah)
\cite{hbar, maser}.
A wide range of recent results
investigating the SME exists and continues to grow.
This includes studies of
neutrinos
\cite{neutrino},
electromagnetism
\cite{km,em},
electromagnetic cavities
\cite{cavity},
clock-comparison experiments
\cite{cc},
satellites
\cite{space},
hydrogen molecules
\cite{hmol},
neutral-meson oscillations
\cite{mesons},
muons
\cite{muons},
baryogenesis
\cite{baryon},
the torsion pendulum
\cite{torsion},
spacetime varying couplings
\cite{coupling},
the Higgs sector
\cite{higgs},
and noncommutative coordinates
\cite{noncommutative}.

\section{Penning Traps}
In general, Penning traps employ a strong uniform magnetic field
to confine charged particles
within a region close to a central symmetry axis.
Particles with one sign of charge
are prevented from drifting in the $B$-field direction by
an electric field with a quadrupole component.
The Penning trap is capable of trapping a single particle for periods of
several months and can be used to make high-precision measurements
of oscillation frequencies of the particle.
The absolute frequency resolutions of Penning traps
make them excellent devices for testing Lorentz symmetry.
In the case of electrons,
Lorentz tests have been performed
based on frequency resolutions of a few Hertz \cite{ptexp}.
Planned experiments with protons and antiprotons
should improve the precision of measurements of
the gyromagnetic ratios of both particles
\cite{ptexp2}.

Since Lorentz-violating effects are minuscule,
they can be calculated in the framework of conventional
perturbation theory.
It turns out that the energy levels of the trap
are dominated by the magnetic field,
so one way to calculate the effects of Lorentz violation
on the energy levels is to use the relativistic Landau levels
of the particle in the uniform $B$ field
as the unperturbed wave functions
for the first-order energy corrections.
For our calculations,
we take a magnetic field aligned with the $z$ axis,
choose the gauge $A^\mu = (0,-yB,0,0)$,
and denote the stationary states
with principal quantum number $n$
and spin $s$
by $\ch^p_{n,s}$ for protons,
and by $\ch^{\overline p}_{n,s}$ for antiprotons.

The SME provides coefficients that account for possible Lorentz
violations in each flavor of fundamental particle.
Some difficulties in treating the constituent quarks and antiquarks
of the proton and antiproton can be overcome by
treating protons and antiprotons themselves as fundamental.
Thus, effective parameters
$a^p_\mu$, $b^p_\mu$, $H^p_{\mu\nu}$, $c^p_{\mu\nu}$, $d^p_{\mu\nu}$
are used.
In the framework of relativistic quantum mechanics,
the perturbation hamiltonian for a proton is found to be
\bea
\hat H_{\rm pert}^p
&=&
a^p_\mu \ga^0 \ga^\mu - b^p_\mu \ga_5 \ga^0 \ga^\mu
- c^p_{0 0} m \ga^0 - i (c^p_{0 j} + c^p_{j 0})D^j
+ i (c^p_{0 0} D_j - c^p_{j k} D^k) \ga^0 \ga^j
\nonumber \\
&& - d^p_{j 0} m \ga_5 \ga^j + i (d^p_{0 j} + d^p_{j 0}) D^j \ga_5
+ i (d^p_{0 0} D_j - d^p_{j k} D^k) \ga^0 \ga_5 \ga^j
+ \half H^p_{\mu \nu} \ga^0 \si^{\mu \nu}.
\nonumber \\
\quad
\label{Hint}
\eea
Certain coefficients for Lorentz violation
can be eliminated by field redefinitions
\cite{fieldredef}.
For the antiproton,
the perturbation hamiltonian
$\hat H_{\rm pert}^{\bar p}$
differs in several negative signs that appear
in the charge-conjugation process.
Thus, the corrections to the Landau energy levels are found
from
\beq
\de E_{n,s}^{p} = \int
\ch_{n,s}^{p \dagger} \, \hat H_{\rm pert}^{p} \,
\ch_{n,s}^{p} \, d^3r
\quad , \qquad
\de E_{n,s}^{\bar p} = \int
\ch_{n,s}^{\bar p \dagger} \, \hat H_{\rm pert}^{\bar p} \,
\ch_{n,s}^{\bar p} \, d^3r
\quad .
\label{delE}
\eeq

The leading-order corrections to the proton energy levels
are
\bea
\de E_{n,\pm 1}^{p} &\approx& a_0^p \mp b_3^p
- c_{00}^p m_p \pm d_{30}^p m_p \pm H_{12}^p
\nonumber \\ &&
- \half (c_{00}^p + c_{11}^p +c_{22}^p) (2n + 1 \mp 1) \om_c
\nonumber \\ &&
- \left( \half c_{00}^p + c_{33}^p \mp d_{03}^p \mp d_{30}^p
\right) \fr{p_z^2}{m_p}
\quad ,
\label{prodelE}
\eea
and these produce shifts in the measured trap frequencies.

One of these frequencies is the cyclotron frequency,
defined in the unperturbed case by
\beq
\om_c = E_{1,+1}^{p} - E_{0,+1}^{p}
\quad ,
\eeq
where $E_{n,s}^p$ are the unperturbed energies of the proton.
Another is the Larmor frequency, defined by
\beq
\om_L = E_{0,-1}^{p} - E_{0,+1}^{p}
\quad .
\eeq
Using equation
(\ref{prodelE}),
and the corresponding antiproton energy shifts,
the frequencies as perturbed by Lorentz violation are
\bea
\om_c^{p} &=& \om_c^{\bar p} \approx
(1 - c_{00}^p - c_{11}^p - c_{22}^p) \om_c
\quad ,
\label{wcp}
\\
\om_L^{p} &\approx& \om_L
+ 2 b_3^p - 2 d_{30}^p m_p - 2 H_{12}^p
\quad ,
\label{wLp}
\\
\om_L^{\bar p} &\approx& \om_L
- 2 b_3^p - 2 d_{30}^p m_p - 2 H_{12}^p
\quad ,
\label{wLpbar}
\eea
at leading order.

In the expressions above,
the subscripts on the
coefficients of Lorentz violation
refer to coordinates fixed in the laboratory reference frame.
As an experiment runs over a period of months,
the laboratory rotates relative to the fixed
background stars.
It also moves
at about 8 kilometers per second around the center of the Earth
and at about 30 kilometers per second around the Sun.
The SME coefficients for Lorentz violation
are fixed
in the Sun-based standard inertial reference frame
that has been adopted for tests of the SME,
whereas the measured laboratory observables
are time dependent.
For any Earth-fixed laboratory,
this dependence includes the sidereal period
of just under 24 hours.
In principle,
other periods can be introduced by rotating turntables
as has been done for example
in some cavity experiments \cite{cavity}.

To compare results from different experiments
it is useful to express the bounds
in the standard inertial-frame coordinates $(T,X,Y,Z)$.
The transformations between the laboratory frame coordinates
$(t,x,y,z)$ and the inertial ones
are discussed in Appendix C of Reference \cite{km}.

Experimental detection of Lorentz violation
in this type of experiment
involves the comparison of two frequencies,
and can be put into two broad categories:
{\em sidereal} and {\em instantaneous}.

Sidereal tests look for time dependence in one frequency
by comparing it with a stable frequency reference.
This could be done, for example,
with a trapped proton by subtracting from the Larmor frequency $\om_L^p$
a fixed reference frequency.
In general,
even the reference frequency will be affected by the coefficients
for Lorentz violation,
as is known from studies of the SME effects
on atomic clocks \cite{cc},
masers \cite{maser},
and microwave cavities \cite{cavity}.
Therefore, the reference clock must be
differently affected by the SME background.
In a sidereal test monitoring the Larmor frequency,
the dependence on the magnetic field is likely to be
a dominant limitation on the precision.
One way to handle this is to use the cyclotron frequency $\om_c^p$
as the reference clock,
since it has the same fractional $B$-field dependence
as the Larmor frequency.
The different SME-coefficient dependence of the two frequencies
could provide a feasible test of Lorentz violation without suppression.
For example,
one could monitor the difference
\beq
\om_L^p - \fr g 2 \om_c^p
\approx
 2 b_3^p - 2 d_{30}^p m_p - 2 H_{12}^p
\quad ,
\label{sidereal:p}
\eeq
which would be zero in the absence of Lorentz violation,
but which is sensitive at leading order to the quantity
$\tilde b _3 = b_3^p -  d_{30}^p m_p -  H_{12}^p$.
If this signal showed no Fourier component at the sidereal frequency
and fell within a one-Hertz error bar
then $\tilde b _3$ would be bounded at the level of $10^{-25}$~GeV.
If a sidereal test was done with a single trapped antiproton,
a similar bound could be envisaged by monitoring the difference
\beq
\om_L^{\bar p} - \fr g 2 \om_c^{\bar p}
\approx
- 2 b_3^p - 2 d_{30}^p m_p - 2 H_{12}^p
\quad ,
\label{sidereal:pbar}
\eeq
thereby measuring a combination of parameters
with opposite sign for  $b_3^p$.

Instantaneous tests involve the comparison
of two frequencies measured at effectively the same time,
so that orientational issues are minimized.
For example,
if the Larmor frequency for a single trapped proton
and a single trapped antiproton
in the same magnetic field could be measured simultaneously,
then the difference in these quantities
\beq
\om_L^p - \om_L^{\bar p}
\approx
4 b_3^p
\quad ,
\label{instant}
\eeq
would be a measure of Lorentz violation.
The advantage of this type of test is the access it gives to
cleaner bounds: instead of bounding a combination of coefficients,
it isolates a single component  $b_3^p$.
This quantity is still time-dependent
because the bound is on the component of $b_j^p$ in the direction of the magnetic field,
which rotates relative to the solar-based inertial reference frame.
Access to such clean bounds is possible
only in the relatively small number of experiments that
compare particles and antiparticles.
For example,
atomic-clock tests cannot do this type of instantaneous test
since this would require antiatom-based clocks.
One could envisage this test performed with two traps
held within the same magnetic field.
Even with one trap,
it should be possible to do an instantaneous test,
since the orientational information could be accounted
for by time-binning the experimental data.
A major limitation would then be the magnetic-field stability,
although this could perhaps be handled by making the comparison
between the quantities in equations (\ref{sidereal:p})
and (\ref{sidereal:pbar}),
rather than between the two Larmor frequencies.

\section{Hydrogen and Antihydrogen}
Several tests of Lorentz symmetry
based on high-precision spectroscopy have been done \cite{maser}.
With the addition of antihydrogen spectroscopy in the future,
it would be possible to perform new tests of Lorentz and CPT symmetry.
A preliminary analysis of the range of possible Lorentz-violating
effects has been completed \cite{hbar},
and in this section some of those results are reviewed.
The status of current efforts to trap antihydrogen atoms
is discussed elsewhere in this conference proceedings.
The discussion below considers spectroscopy with trapped
antihydrogen, based on quantum-mechanical stationary states
modified by the presence of a uniform magnetic field.
An analysis of transition frequencies of free \h\ and \ah\
has been done elsewhere and shows no leading-order effects \cite{hbar}.

The effects of Lorentz violation
on \h\ and \ah\
can be calculated at leading order
using the relativistic solutions of the Dirac
equation for free \h\ and \ah\ as the unperturbed states.
In the case of \h,
the perturbation hamiltonian for the electron
is of the same form as equation (\ref{Hint})
but with the superscripts $p$ replaced by $e$.
The Lorentz-violating effects involving the proton
can also be included without difficulty.
There are four possible spin
configurations for each state in the decoupled basis,
determined by the choice of component of angular momentum
along the quantization axis, $m_J=\pm 1/2$ and $m_I = \pm 1/2$,
where $J$ and $I$ are the electron and proton angular momenta.
The shifts in the energy levels
are found to be:
\bea
\De E^{H} (m_J, m_I)
& \approx &
(a_0^e + a_0^p - c_{00}^e m_e - c_{00}^p m_p)
\cr
&&
+ (-b_3^e + d_{30}^e m_e + H_{12}^e) {m_J}/{|m_J|}
\cr
&&
+ (-b_3^p + d_{30}^p m_p + H_{12}^p) {m_I}/{|m_I|} \quad .
\label{EHJI}
\eea

For trapped \h\ and \ah,
we assume a uniform magnetic field $B$
that splits the 1S and 2S levels into four hyperfine Zeeman levels,
denoted in order of increasing energy by $\ket{a}_n$, $\ket{b}_n$,
$\ket{c}_n$, $\ket{d}_n$,
with principal quantum number $n=1$ or $2$, for both
\h\ and \ah.
The $\ket c$ and $\ket d$ states are trapped
and so it is of interest to consider transitions
involving these states.
The shifts in energies of the states
$\ket{d}_1$ and $\ket{d}_2$
are identical,
so no leading-order effect on this
particular 1S-2S transition is seen
for either \h\ or \ah.

In the case of the
1S-2S transition between the
states $\ket{c}_1$ and $\ket{c}_2$ in \hydrogen\ and \ah,
an unsuppressed frequency shift does occur.
This is because the dependence on $n$ in the hyperfine splitting
produces a spin-mixing difference
between the 1S and 2S levels.
The effect can be shown to be optimum
at a magnetic field of about $B \simeq 0.01$ Tesla
for both \h\ and \ah.
The magnetic field gradient of the trap
may be a limitation in this case.

The hyperfine transitions within a single energy level
are also of interest for Lorentz tests in \h\ and \ah.
It is found that the coefficients for Lorentz violation
give rise to field-dependent energy shifts
of the $\ket a$ and $\ket c$ states
and field-independent shifts of
the $\ket b$ and $\ket d$ states.
In the case of the
$\ket{d}_1 \longrightarrow \ket{c}_1$ transition,
a field-independent point exists at about 0.65~Tesla,
which may be useful experimentally.
At this field value,
the transition is essentially a proton spin flip
and so the coefficients for Lorentz violation
that are bounded are those of protons.
The relevant leading-order shifts
in the frequencies $\nu_{c \rightarrow d}^H$ and
$\nu_{c \rightarrow d}^{\overline{H}}$
for \h\ and \ah\ respectively are:
\bea
\de \nu_{c \rightarrow d}^H
&\approx&
(-b_3^p + d_{30}^p m_p + H_{12}^p)/\pi
\quad , \label{freq:h:cd}\\
\de \nu_{c \rightarrow d}^{\overline{H}}
&\approx&
(b_3^p + d_{30}^p m_p + H_{12}^p)/\pi
\quad . \label{freq:Hbar:cd}
\eea
Expressions
(\ref{freq:h:cd}) and (\ref{freq:Hbar:cd})
are similar to the expressions (\ref{wLp}) and (\ref{wLpbar})
discussed for the Penning trap
and this is because both involve the spin-flip transition
of a proton in a magnetic field.

Sidereal tests looking for Lorentz violation
in conventional \h\ have already been conducted \cite{maser}
at the Harvard-Smithsonian Center for Astrophysics,
based on the  $F=1$, $\De m_F=\pm 1$ transition.
Since a weak field was used,
the bound is on a mixture of electron and proton parameters
of the form seen in equation (\ref{freq:h:cd}).
A frequency stability of about one mHz
gives an upper bound at the level of about $10^{-27}$~GeV.
Sidereal tests with \ah\ are not yet possible,
but in principle could be done by monitoring
the hyperfine transition and comparing it to a stable clock.
The bound would be on the Lorentz-violation combination
seen in equation (\ref{freq:Hbar:cd}).

Instantaneous tests comparing the hyperfine transitions
in \h\ and \ah\ would bound the difference between the two
quantities in
(\ref{freq:h:cd}) and (\ref{freq:Hbar:cd}):
\beq
\De \nu_{c \rightarrow d} \equiv
\nu_{c \rightarrow d}^H - \nu_{c \rightarrow d}^{\overline{H}}
\approx - 2 b_3^p / \pi
\quad .
\eeq
The advantage of this type of test is the clean bound
obtained on just one coefficient for Lorentz violation.
A non-zero value of $b_3^p$ would indicate CPT violation,
whereas the combination of coefficients appearing
would only indicate Lorentz violation.
If a 10-mHz resolution was obtained
the bound would be at the level of $10^{-26}$~GeV.

\section{Discussion}
The sensitivity of the Lorentz tests discussed here
is determined by absolute frequency resolutions $\de\nu$
and not by the relative precisions $\de\nu / \nu$.
To optimize such tests one may be able to select
experimental variables so as to improve the absolute
resolution,
even when this seems counterintuitive
since no gain appears in relative precision.
In the case of the Penning trap,
if the magnetic field could be reduced by a factor of $10$,
without reduction in the relative precision,
the bound on the SME coefficients would improve tenfold
because of the reduction in the Larmor frequency.

The SME is a useful tool for seeking promising tests of Lorentz violation.
One significant feature is the ability to compare results of Lorentz tests
in different experiments.
In particular,
the Penning trap experiments with protons and antiprotons,
and spectroscopy experiments with \h\ and with \ah\ have a
significant area of overlap.
However,
since the coefficients for Lorentz violation
are those for protons and electrons only,
there is no overlap with CPT tests with kaons, for example.
The SME also provides a means of identifying unsuppressed transitions.
For example,
in the case of free \h\ and \ah,
effects in the 1S-2S transition are suppressed by two powers of the fine-structure constant.

It is also worth noting that most of the Penning-trap discussion
for protons and antiprotons
holds also for similar experiments
with electrons and positrons,
using the anomaly frequency in place of the Larmor frequency.
Bounds from several such experiments exist
\cite{ptexp},
based on experimental data taken more than twenty years ago
and could possibly be improved with present technology.

\bibliography{sample}

\end{document}